\documentclass[letterpaper]{article}
\pdfoutput=1 
\usepackage{mume}
\usepackage{times}
\usepackage{helvet}
\usepackage{courier}

\begin{document}
	%

	\title{Generating Albums with SampleRNN\\to Imitate Metal, Rock, and Punk Bands}

	\author{CJ Carr \\
		Dadabots \\
		emperorcj@gmail.com
      \And
	    Zack Zukowski \\
		Dadabots \\
		thedadabot@gmail.com
	}
	\maketitle

	\begin{abstract}
		\begin{quote}

	This early example of neural synthesis is a proof-of-concept for how machine learning can drive new types of music software. Creating music can be as simple as specifying a set of music influences on which a model trains. We demonstrate a method for generating albums that imitate bands in experimental music genres previously unrealized by traditional synthesis techniques (e.g. additive, subtractive, FM, granular, concatenative). Raw audio is generated autoregressively in the time-domain using an unconditional SampleRNN. We create six albums this way. Artwork and song titles are also generated using materials from the original artists' back catalog as training data. We try a fully-automated method and a human-curated method. We discuss its potential for machine-assisted production.

		\end{quote}
	\end{abstract}
	
	\section{Background }

    While symbolic approaches to generating music are becoming increasingly capable of capturing subtle time and dynamic variations in a performance \cite{performanceRNN}, musical features such as timbre and space are not easily represented in MIDI. 
        
    Since the twentieth century, the study of manipulating timbre has played a much more significant role in composition technique. Composers like Varese thought in terms of composing ``sound-masses" to construct his symphonic scores \cite{varese} inspiring future artists to discover new formal organizational structures involving an extended pallet of sonic material.

	The majority of deep learning papers on generative music focus on symbolic-domain generation, including all seven that appeared at ISMIR 2017 \cite{7,8,9,10,11,12,13}. Few have explored recent advances in neural synthesis of raw audio such as Wavenets \cite{wavenet}, SampleRNN \cite{Mehri2017}, DeepVoice \cite{deepvoice}, TacoTron 2 \cite{tacotron2}, and WaveRNN \cite{waveRNN}. Those that have explored neural synthesis have focused on piano \cite{Vasanth} or classical music \cite{Mehri2017}.
    
    Most style-specific generative music experiments have explored artists commonly found in harmony textbooks such as Bach \cite{13} and Beethoven \cite{Mehri2017} but few have looked at generating modern sub-genres with subtle stylistic distinctions such as black metal, math rock, and skate punk. Perceiving the nuanced spectral characteristics found in modern production can be difficult for untrained human listeners to describe and are poorly represented by traditional transcriptions of music, thus they require a raw audio generative model to reproduce. 
    
    Earlier work \cite{dadabots1} details our initial experiments with SampleRNN which evaluated its music synthesis capabilities in an extended range of modern musical styles.

	\section{Method }

    \subsection{SampleRNN}
    
    SampleRNN is a recurrent neural network. Recurrent neural networks are used for sequence prediction \cite{karpathy,graves,schmidhuber}, i.e. given what has happened previously in a sequence, what will happen next? Music can be modeled as a sequence of events over time. Thus, music can be generated by predicting ``and then what happens?"  again and again. This technique is called autoregressive generation. In audio, it is one example of neural synthesis.
    
    SampleRNN was originally developed for text-to-speech applications. By training it on examples of raw audio speech, it learns to generate new raw audio examples of speech. If the model is trained on metadata (e.g. phonemes, speaker id) at the same time as its associated raw audio, it is called ``conditional" if as a result the generated audio can be controlled or driven to make specific speech patterns. For example, using a conditional speech model, we can make it say ``Sorry Dave, I'm afraid I can't do that" even if it has never seen that phrase before in the training data. 
    
    An unconditional model does not have this property. It is not controlled or driven during autoregressive generation. Instead it wanders through its state space, conditioned only on the state of the previous timestep. We argue that this gives them a distinct sonic behavior. 
    
    SampleRNN predicts what audio sample comes next, one sample at a time. Depending on the sample rate of the audio, this could mean 16000, 32000 or 44100 predictions per second. During training, the gradient from this prediction loss backpropagates through time. In a sense, it updates the choices it should have made in order to guess correctly. This backpropagation is truncated to a few dozen milliseconds, however, because of limited computational resources. Nevertheless, we observe that the output of a trained model is rather musical, invents riffs and melodies, and transitions between sections. 
    
    The goal is to synthesize music within the limited aesthetic space of the album's sound. Thus, we want the output to overfit short timescale patterns (timbres, instruments, singers, percussion) and underfit long timescale patterns (rhythms, riffs, sections, transitions, compositions) so that it sounds like a recording of the original musicians playing new musical compositions in their style. 

    Because of unconditional SampleRNN's samplewise loss function and truncated backpropagation through time, it primarily focuses on short timescale patterns, gradually modeling longer timescales the longer it trains.  Whereas a top-down progressive learning method \cite{progressiveGAN} would do the opposite.
    
    We forked the SampleRNN theano code on github and adapted it for music. We ran several dozen experiments to find hyperparameters and dataset choices which worked well for generating music. 
   
    Our github fork can be found here at this link https://github.com/ZVK/sampleRNN\_ICLR2017
   
	\subsection{Music Dataset}
	
    We prepare an audio dataset for SampleRNN. The choice of audio which goes into the dataset is significant. We  observe the effects of this choice on the generated audio (e.g. how musical does it sound, how close is it to the original music?) though we have no formal method of measuring this. 
    
    We observe that the model generates better audio if the dataset has varied music patterns but consistent instrumentation, production, and mastering characteristics. 
    
    Therefore, for each dataset, we choose to use one album from one artist, because it represents the longest cohesively packaged musical idea. Teams of producers and audio engineers ensure that a commercial album is normalized, dynamically compressed, denoised, and filtered to deliver a consistent aural quality across various listening environments. 
        
    Sometimes a trained model generates output with long silences. To mitigate this, we preprocess the audio dataset to remove silences.

    \subsection{Generating Music}

    For each generated album, we generate music, album artwork, and song titles learned from small datasets deriving from subsets of a single artist's discography. A large batch of content is generated. In the fully-automated method, content is chosen randomly from this batch. In the human-curated method, a curator chooses from this batch. 
    
    We split each audio dataset into 3,200 eight-second overlapping chunks of raw audio data (FLAC). The chunks are randomly shuffled and split into training, testing, and validation sets. The split is 88\% training, 6\% testing, 6\% validation. We train each model for two days on a NVIDIA V100 GPU on AWS.

    We use a 2-tier SampleRNN with 256 embedding size, 1024 dimensions, 5 layers, LSTM, 256 linear quantization levels, 16kHz or 32kHz sample rate, skip connections, and a 128 batch size, using weight normalization. The initial state h0 is randomized to generate more variety.

    Intermittently at checkpoints during training, audio clips are generated autoregressively, one sample at a time, and converted to WAV files. About ten hours of music is generated.

	\subsection{Generating Artwork}

We use a few methods to algorithmically generate album art. One method is to run neural style transfer \cite{styletransfer} on the original album cover, using the same image as both content and style, with the style downscaled, resulting in an effect which looks like the cover were made out of small pieces of itself. 

Another method is to style transfer the album cover with photographs of the band taken from their press kit or social media feed, generate multiple style transfers from different photographs, then overlay them together with an overlay filter. 

\subsection{Generating Titles}

We take the artist's entire discography of song titles, run them through a second- and third-order Markov chain, and generate thousands of new titles. In the fully-automated system, random titles are chosen for songs. In the human-curated system, the curator chooses titles and matches them with generated songs. 

\section{Results}

We ran initial experiments on multiple genres including electronic, hip-hop, black metal, mathcore, rock, and skate punk. 

We observed that electronic music and hip-hop instrumentals did not seem to train as well as organic, electro-acoustic ensembles of musicians. Music genres like metal and punk seem to work better, perhaps because the strange artifacts of neural synthesis (noise, chaos, grotesque mutations of voice) are aesthetically pleasing in these styles. Furthermore, their fast tempos and creative use of loose performance techniques translate well to SampleRNN's rhythmic distortions. 

As we experimented, we tweaked our process, introducing varying degrees of human curation. We took what we felt were the best results and turned them into albums. Albums can be heard at http://dadabots.com or http://dadabots.bandcamp.com

    \begin{itemize}
		\item 	\textbf{ Deep The Beatles}
    ---  Trained on ``ONE" by The Beatles. Their greatest hits. We play the listener a random 30-second audio clip from each training epoch. This reveals the learning process of the machine. At first the clips are noisy and textural. Then voice and percussion enter. It never learns rhythm or song structure though. The training is cut off at 22 epochs. 
\item \textbf{ Calculating Calculating Infinity}
    ---  Trained on ``Calculating Infinity" by The Dillinger Escape Plan. Mathcore with screams and spastic time signatures. Random 4-minute audio clips, generated from the final epoch, were chosen  for the tracks. There was no human curation or consideration of album flow. The autoregression occasionally got trapped in a section without escaping. Several songs ended up trapped in the same section, like a chaotic attractor, making this album too repetitive at times.
\item \textbf{ Coditany of Timeness}
    ---  Trained on ``Diotima" by Krallice. Black metal with atmospheric texture and tremolo-picked guitars. This was the first model of ours that learned to keep time, heard in the consistent pulse of the blast beats. We listened through all the audio generated in its final epoch, chose a few 1-4 minute sections that sounded like songs with progression and transitions, and arranged them while considering album flow. 
\item \textbf{ Inorganimate}
    --- Trained on ``Nothing" by Meshuggah. Math metal with 8-string guitars and complex polyrhythms. We continued the trend of human curation, this time stitching sections from various epochs together to make songs with the full variety of generated sound. The earlier epochs had weird texture and percussion effects. The middle epochs came up with tempos and vocal styles not used in the album. The later epochs better replicated the band's palette but arhythmically and awkwardly (though the effect is humorous). 
\item \textbf{ Megaturing}
    ---  Trained on ``Mirrored" by Battles. Experimental rock with rigid drum grooves and electro-acoustic polyrhythmic textures. We extended the human curated approach by introducing a new audio layering technique intended to create a stereo image from monophonically generated samples. The accent beats of these layered sections were temporally aligned and allowed to drift naturally when harmonizing.
\item \textbf{ Bot Prownies}
    ---  Trained on ``Punk In Drublic" by NOFX. Skate punk with melodic vocals. Some of the songs we left unedited (e.g. ``Bot Prownies", ``One Million Governments") as we discovered them in the batch. Some of the songs (e.g. ``Jong Out", ``Lose Home") we edited by repeating sections and motifs, increasing their musicality. We picked titles from the Markov chain output that meaningfully fit the songs. The lyrics were nonsensical syllables, as the model does not learn a language model. Nevertheless two fan-made ``misheard lyrics" music videos were created, where the fans attempted to interpret the lyrics.
	\end{itemize}

We do not yet have a formal way of measuring the quality of generated audio nor its closeness to the original music. This would be not only interesting but necessary for the development of better loss functions which consider the psychoacoustic perception of music and music style. 

We observe that the fully-automated method, with no consideration of album flow, produced albums that were at times repetitive and challenging for the listener to engage. Whereas introducing human curation increased listenability and variety. 

Curating with neural synthesis is not unlike working with a recorded jam session, listening through an improvisation, picking the best moments, and weaving them into an album. 

While we set out to render convincing likeness of these bands, we were delighted by the aesthetic merit of the imperfections. Pioneering artists can exploit these effects, just as they exploit vintage sound production (tube warmth, tape-hiss, vinyl distortion, etc).

Though it is impressive to generate fully-automated albums without human intervention, we would like to emphasize human-centric design \cite{fiebrink} and mixed-initiative generation \cite{mixed}. How can we enhance the workflows of artists? What new forms of collaboration are possible? How will neural synthesis change digital audio workstations (e.g. Ableton Live, FL Studio)?

We believe these early examples of neural synthesis are proofs-of-concept for how machine learning will drive new types of music software, granting their users greater high-level control and accessibility. We are confident that real-time conditional generation will enable new expressive digital music instruments. The limiting factor thus far is computational efficiency, but this can be solved with better algorithms (parallelism, pruning, etc) and faster hardware.

\section{Conclusion }

In this paper, we demonstrated that creating music can be as simple as specifying a set of music influences on which a machine learning model trains. We applied our method on different music genres (e.g. metal, punk) previously unrealized by traditional synthesis techniques (e.g. additive, subtractive, FM, granular, concatenative). Raw audio was generated autoregressively in the time-domain using an unconditional SampleRNN. Six albums were generated including music, artwork, and song titles. We explored fully-automated and human-curated methods, and observed human-curation increased album listenability and variety. We discussed the potential for machine-assisted production and live digital instruments.

\section{Future Work}

\subsection{Future Technical Work}

Future technical work includes (1) local conditioning with hybrid representations of raw and symbolic audio, (2) exploring Dilated RNN \cite{dilatedRNN}, Nested LSTM \cite{nestedLSTM}, (3) style transfer \cite{audioStyleTransfer}, (4) generating stereo and multitrack recordings, (5) conditioning sample-level tiers on spectrogram-level tiers \cite{tacotron2}, (6)  progressively-trained generative models \cite{progressiveGAN} that focus on song structure, (7)  student networks \cite{parallelWavenet}, weight pruning \cite{waveRNN}, and parallelization to increase generative speed to real-time, (8)  formal methods of measuring the quality of generated music, (9)  improved loss functions, and (10) a new type of digital audio workstation based on neural synthesis.

\subsection{Future Artistic Work}

Imminent collaborative work we are doing with artists include building a generative model for UK champion beatboxer Reeps One so that he can battle himself as part of a Bell Labs documentary.

\bibliographystyle{mume}
\bibliography{main}

\end{document}